\documentclass[journal]{IEEEtran}
\usepackage{mathrsfs}
\usepackage{tikz}
\usepackage{color}
\usepackage{amsmath}
\usepackage{amssymb}
\usepackage{amsthm}
\usepackage{graphicx}
\usepackage{extarrows}
\usepackage{times}
\usepackage{graphics,color}

\newcommand{\oneton}{1,\cdots,n}

\graphicspath{{figs/}}

\newtheorem{mythm}{Theorem}
\newtheorem{mycol}{Corollary}

\newtheorem{myrem}{Remark}
\begin{document}
\title{
Another Approach to Consensus of Multi-agents}

\author{Tianping~Chen,~\IEEEmembership{Senior~Member,~IEEE}
\thanks{This work is  supported by the National Natural Sciences Foundation of China under Grant Nos. 61273211.}
\thanks{T. Chen was with the School of Computer
Sciences/Mathematics, Fudan University, Shanghai 200433, China (tchen@fudan.edu.cn).
}}

\maketitle

\begin{abstract}
In this short note, we recommend another approach to deal with the topic Consensus of Multi-agents, which was proposed in \cite{Chena}.

\end{abstract}



In \cite{Lu2006}, following model was discussed
\begin{align}\label{model1}
\frac{d x_{i}(t)}{dt}=f(x_{i}(t),t)
+c\sum\limits_{j=1}^{m}l_{ij}\Gamma x_{j}(t),\quad i=1,\cdots,m
\end{align}
where $x_{i}(t)\in R^{n}$ is the state variable of the $i-th$
node, $t\in [0,+\infty)$ is a continuous time,
$f:R\times[0,+\infty)\rightarrow R^{n}$ is continuous map,
$L=(l_{ij})\in R^{m\times m}$ is the coupling matrix with zero-sum
rows and $l_{ij}\ge 0$, for  $i\ne j$,  which is determined by the
topological structure of the LCODEs, and $\Gamma\in R^{n\times n}$ is an inner coupling matrix. Some time, picking
$\Gamma=diag\{\gamma_{1},\gamma_{2},\cdots,\gamma_{n}\}$ with
$\gamma_{i}\ge 0$, for $i=\oneton$.

\begin{align}\label{model2}
\frac{d x^{i}(t)}{dt}=Ax^{i}(t)
+c\sum\limits_{j=1}^{m}l_{ij}\Gamma x^{j}(t),\quad i=1,\cdots,m
\end{align}
where $A\in R^{n\times n}$.

In case that the state variables $x_{i}(t)$ are not observed. Then, instead of coupling $x_{i}(t)$ (because they are not available),  in \cite{Li1} and some other papers, the authors coupled the measured output
\begin{align*}
\dot{\zeta}_{i}(t)=\sum_{j=1}^{N}l_{ij}y_{i}(t)
\end{align*}
and following observer based synchronization model
\begin{align}\label{model4}
\frac{d x_{i}(t)}{dt}=Ax_{i}(t)
+c\sum\limits_{j=1}^{m}l_{ij}FC x_{j}(t),\quad i=1,\cdots,m
\end{align}
is proposed, where $y(t)=Cx(t)$ is observer measurement $C\in R^{q\times n}$, and $C\in R^{n\times q}$,  was discussed.

It is clear that all these models are special cases of the most general and universal model (\ref{model1}).

In the following, we investigate the model  
\begin{align}
\frac{d x_{i}(t)}{dt}=Ax_{i}(t)
+c\sum\limits_{j=1}^{m}l_{ij}\Gamma x_{j}(t),\quad i=1,\cdots,N
\end{align}
with another approach proposed in \cite{Chena}.

Firstly, we recall the results given in \cite{Chena}.

Denote
$y_{i}(t)=x_{i}(t)-x_{1}(t)$,
$i=2,\ldots,m$. Then,
$y_{i}(t)=Rx_{i}(t)$, where
\begin{eqnarray*} R=\left[\begin{array}{ccccc}
        -1       & 1  & \cdots & 0     \\
         \cdots  & \cdots & \ddots & \cdots\\
        -1       & 0 & \cdots     & 1
  \end{array}\right]_{(m-1)\times m}
\end{eqnarray*}

It is easy to see that the Moore-Penrose inverse of $R$ can be
written as \setlength{\arraycolsep}{1pt}
\begin{eqnarray*}
&   &
R^{\dag}=\\
&   &
\setlength{\arraycolsep}{5pt}
~\frac{1}{m}
\left[\begin{array}{ccccc}
             -1   & -1    & \cdots & -1 &-1\\
             (m-1)    & -1  & \cdots & -1&-1\\
             \cdots    & \cdots & \ddots & \cdots & \cdots\\
             -1    & -1    & \cdots & (m-1)&-1\\
             -1    & -1   & \cdots & -1&(m-1)
       \end{array}\right]_{m\times (m-1)}
\end{eqnarray*}
\setlength{\arraycolsep}{5pt}

Denote $RLR^{\dag}=L^{*} $. Since $L$
is a zero-row-sum matrix, we have
\begin{eqnarray*}
L^{*}=\left[\begin{array}{ccccc}
        l_{22}-l_{12}  & \cdots & l_{2m}-l_{1m}\\
        \cdots & \ddots & \cdots\\
         l_{m2}-l_{12} &\cdots & l_{mm}-l_{1m}\\
       \end{array}\right]_{(m-1)\times (m-1)}
\end{eqnarray*}
Then, we have
\begin{align}
\dot{y}_{i}(t)=Ay_{i}(t)+c\sum_{j=2}^{m}L^{*}_{ij}\Gamma y_{j}(t),~i=2,\cdots,N
\end{align}

Furthermore, let $\lambda_{1},\lambda_{2},\cdots,\lambda_{m}$ be the eigenvalues of $L$ with $\lambda_{1}=0$. Then $\lambda_{2},\cdots,\lambda_{m}$ be the eigenvalue decomposition of the matrix $L^{*}$.

Based on these observations given in \cite{Chena}, we can easily deal with consensus of multi-agents

Let $L^{*}=Q^{-1}\Lambda_{1} Q$ be its eigen-decomposition, where
$\Lambda_{1}=\mathrm{diag}\{\lambda_{2},\cdots,\lambda_{m}\}$, and
$z_{k}(t)=Qy_{k}(t)$.
\begin{align}
\dot{z}_{i}(t)=Az_{i}(t)+c\lambda_{i}\Gamma z_{i}(t)
\end{align}
It is clear that $x_{i}(t)$ reaches consensus is equivalent to all $z_{i}(t)$, $i=2,\cdots,m$ converge to zero.

Therefore, we have
\begin{mythm}\quad
Let $\lambda_{2},\lambda_{3},\cdots,\lambda_{m}$ be the non-zero
eigenvalues of the coupling matrix $L$. If all variational
equations
\begin{align}
\dot{u}(t)=[A+c\lambda_{i}\Gamma] u(t),~~k=2,3,\cdots,m
\end{align}
are exponentially stable, then the consensus of model (\ref{model2}) can be reached
exponentially for the coupled system.
\end{mythm}

\begin{mythm}\quad Let $\lambda_{k}=\alpha_{k}+j\beta_{k}$, $k=2,\cdots,m$, where $j$
is the imaginary unit, be the eigenvalues of the coupling matrix. If there exist a positive definite matrix $P$ and $\epsilon>0$ such that
\begin{eqnarray}
\bigg\{P(A+c\lambda_{k}\Gamma)\bigg\}^{s}<-\epsilon E_{n},\quad
k=2,3,\cdots,m \label{model1a}
\end{eqnarray}
where $H^{s}=(H^{*}+H)/2$, $H^{*}$ is Hermite
conjugate of $H$, and $E_{n}\in r^{n\times n}$ is identity matrix,
then the consensus of model (\ref{model2}) can be reached
exponentially for the coupled system.

Additionally, if $P\Gamma$ is symmetric and positive definite, then condition (\ref{model1a}) can be replaced by following condition
\begin{eqnarray}
PA+A^{T}P+c Re\{\lambda_{2}\}P\Gamma<-\epsilon E_{n},\quad
\label{model1aa}
\end{eqnarray}
where $Re\{\lambda_{2}\}<0$ is the real part of $\lambda_{2}$.
\end{mythm}

As direct consequences, we have

\begin{mycol}
Let $\lambda_{2},\lambda_{3},\cdots,\lambda_{m}$ be the non-zero
eigenvalues of the coupling matrix $L$. If all variational
equations
\begin{align}
\frac{d z(t)}{dt}=[A+c\lambda_{k}FC]z(t),\quad
k=2,3,\cdots,m
\end{align}
are exponentially stable, then the model (\ref{model4}) can reach consensus exponentially.
\end{mycol}

\begin{mycol}
Let $\lambda_{2},\lambda_{3},\cdots,\lambda_{m}$ be the non-zero
eigenvalues of the coupling matrix $L$. If there exist a positive definite matrix $P$ and $\epsilon>0$ such that
\begin{align}
\bigg\{P(A+c\lambda_{k}FC)\bigg\}^{s}<-\epsilon E_{n},\quad
k=2,3,\cdots,m \label{model1b}
\end{align}
are exponentially stable, then the model (\ref{model4}) can reach consensus exponentially.

In case $(A,C)$ is detectable, then by \cite{Li1} or \cite{Chen6}, we have
$$PA+A^{T}P^{T}-C^{T}C<-\epsilon E_{n}$$
In this case, pick $F=P^{-1}C^{T}$, and $c Re\{\lambda_{2}\}<-1$, then condition (\ref{model1b}) is satisfied.
\end{mycol}

\begin{myrem}
In \cite{Lu2006}, the reference state in synchronization manifold is $\bar{X}(t)=[\bar{x}^{T}(t),\cdots,\bar{x}^{T}(t)]^{T}$, where  $\bar{x}(t)=\sum_{i=1}^{N}\xi_{i}x_{i}(t)$. Any $x=(x^{\top}_{1},\cdots,x^{\top}_{m})^{\top}\in R^{mn}$,
can be written as $x=\bar{X}+\delta x$, and it holds that $\bar{X}\in\mathcal S$ and
$\delta x\in\mathcal L$.

Instead, here, the reference state in synchronization manifold is $X_{1}(t)=[x_{1}^{T}(t)(t),\cdots,x_{1}^{T}(t)(t)]^{T}$. Of course, $x_{1}(t)$ can be replaced by any $x_{i}(t)$, $i=1,\cdots,N$.

It must be noted that in some papers, by letting $x_{1}(t)=s(t)$ with $\dot{s}(t)=f(s(t))$ for the model (\ref{model1}). It is incorrect. In this case the model becomes a master-slave system.
\end{myrem}

Now, based on previous reasoning, we also can use Lyapunov function approach.

Because all eigenvalues of $L^{*}$ are negative, there exists a positive definite matrix $Q$, such that $(QL^{*})^{s}$ is negative definite.

Define Lyapunov function
\begin{align}
V(t)=y^{T}(t)\{Q\otimes P\}y(t) 
\end{align}
\begin{align}
\dot{V}(t)&=y^{T}(t)\{Q\otimes (PA)^{s}\}y(t)
\nonumber\\&+y^{T}(t)\{c(QL^{*})^{s}\otimes P\Gamma\}y(t)
\end{align}

Let $\mu_{1}\ge \mu_{2}\ge \cdots,\mu_{m}>0$ are eigenvalues of the matrix $Q$. $\nu_{1}\ge \nu_{2}\ge \cdots,\nu_{n}$ are eigenvalues of the matrix $(PA)^{s}$. Then
\begin{align}
y^{T}(t)\{Q\otimes (PA)^{s}\}y(t)\le c_{1}y^{T}(t)y(t)
\end{align}
where
$$c_{1}=\max_{k=1,\cdots,m,j=1,\cdots,n}\{\mu_{k}\nu_{j}\}$$

Let $0>\gamma_{2}\ge\cdots\ge\gamma_m$ be the eigenvalues of the matrix $(QL^{*})^{s}$. $\theta_{1}\ge \cdots\ge \theta_{n}>0$ are eigenvalues of $P\Gamma$. Then
\begin{align}
y^{T}(t)\{c(QL^{*})^{s}\otimes P\Gamma\}y(t)\le cc_{2}y^{T}(t)y(t)
\end{align}
where
$$c_{2}=\gamma_2\theta_{n}<0$$

Pick $c$ sufficient large such that $cc_{2}<c_{1}$. Then, we have
\begin{align}
\dot{V}(t)<(cc_{2}+c_{1})y^{T}(t)y(t)
\end{align}

\begin{mythm}\quad
Suppose that $P$ and $P\Gamma$ are positive definite matrices, then for sufficient large coupling strength $c$, then the consensus of model (\ref{model2}) can be reached
exponentially for the coupled system.
\end{mythm}

{\bf Conclusions}
In this short note, by using the results given in \cite{Chena}, we provide an effective approach to deal with consensus of multi-agents.

\end{document}